\newif\iflongversion
\newif\iftwocol

\longversiontrue 

\iflongversion
\twocolfalse
\else
\twocoltrue
\fi

\iftwocol
\documentclass[conference]{IEEEtran}
\else
\documentclass[conference,onecolumn,draftcls,12pt]{IEEEtran}
\fi

\usepackage{header}

\usepackage{kolor}

\usepackage{kontext}

\usepackage{konfig}

\allowdisplaybreaks

\usepackage[hidelinks]{hyperref}

\usepackage{subcaption}

\usepackage{float}
\usepackage{enumitem}
\usepackage{microtype}

\begin{document}

\IEEEoverridecommandlockouts

\title{Nonasymptotic coding-rate bounds for binary erasure channels with feedback}
\author{\IEEEauthorblockN{Rahul Devassy$^1$, Giuseppe Durisi$^1$,  Benjamin Lindqvist$^1$, Wei Yang$^2$, Marco Dalai$^3$}
\IEEEauthorblockA{
 $^1$Chalmers University of Technology, 41296 Gothenburg, Sweden;\\
 $^2$Princeton University, Princeton, NJ, 08544, USA;\ $^3$University of Brescia, 25123 Brescia, Italy\ignorespaces
}
\thanks{This work was partly funded by the Swedish Research Council under grant 2012-4571.
The simulations were performed in part on resources provided by the Swedish National Infrastructure for Computing (SNIC)  at C3SE.}}

\maketitle

\begin{abstract}
We present nonasymptotic achievability and converse bounds on the maximum coding rate (for a fixed average error probability and a fixed average blocklength) of variable-length full-feedback (VLF)  and variable-length stop-feedback (VLSF) codes operating over a binary erasure channel (BEC). For the VLF setup, the achievability bound relies on a scheme that maps each message onto a variable-length Huffman codeword and then repeats each bit of the codeword until it is received correctly. The converse bound is inspired by the meta-converse framework by Polyanskiy, Poor, and Verd\'u (2010) and relies on binary sequential hypothesis testing. For the case of zero error probability, our achievability and converse bounds match. For the VLSF case, we provide achievability bounds that exploit the following feature of BEC: the decoder can assess the correctness of its estimate by verifying whether the chosen codeword is the only one that is compatible with the erasure pattern. One of these bounds is obtained by analyzing the performance of a variable-length extension of random linear fountain codes. The gap between the VLSF achievability and the VLF converse bound, when number of messages is small, is significant:~$23\%$ for 8 messages on a BEC with erasure probability~$0.5.$ The absence of a tight VLSF converse bound does not allow us to assess whether this gap is fundamental.
\end{abstract}\ignorespaces

\section{Introduction}

In a point-to-point communication system with full feedback, the transmitter has noiseless access to all the previously received symbols. For discrete memoryless channels (DMCs), it turns out that this additional information does not increase capacity when codes of fixed blocklengths are used. Specifically, Shannon~\cite{ZeroErrorCapacityShanon} proved that the capacity with full-feedback fixed-blocklength codes is no larger than the one achievable in the no-feedback case. Dobrushin~\cite{doburshin1962anasymptotic} established a similar result for the reliability function of symmetric DMCs (the general case is, however, open). However, if the use of variable-length codes is permitted, the availability of full feedback turns out to be beneficial. Burnashev~\cite{burnashev1976data} derived the reliability function for the case when full feedback is available and variable-length feedback (VLF) codes are used, for all rates between zero and capacity. He showed that this reliability function is
\begin{IEEEeqnarray}{rCl}
E\parantheses{R}&=& C_1\parantheses{1-\frac{R}{C}},\qquad R\in\parantheses{0,C}.\label{expression_burnashev_reliability}
\end{IEEEeqnarray}
Here, $C$ is the channel capacity and $C_1$ denotes the maximum relative entropy between two arbitrary conditional output distributions. Note that~\eqref{expression_burnashev_reliability} is strictly larger than the reliability function for the no-feedback case. Burnashev's proof relies on the asymptotic analysis of an achievability and a converse bound on the maximum rate obtainable with VLF codes, for a given average blocklength and a fixed average error probability. Yamamoto and Itoh~\cite{YamamotoAchievability1979} gave an alternative proof of Burnashev's achievability bound, which relies on a two-phase scheme: a standard transmission phase where feedback is not used at the transmitter is followed by a confirmation phase where the transmitter uses feedback to confirm/contradict the decision of the receiver. Berlin~\etal~\cite{BerlinSimpleConverse2009} provided a stronger version of Burnashev's converse bound, whose proof parallels the two-phase scheme in~\cite{YamamotoAchievability1979}. A one-phase scheme that achieves~\eqref{expression_burnashev_reliability} was proposed in~\cite{JensenShannonDivergenceNaghshvar2015}.

Polyanskiy~\etal~\cite{polyanskiy2011feedback} obtained a nonasymptotic converse bound that improves on Burnashev's one~\cite{burnashev1976data}. In the same work, an achievability bound is provided, which is used to show that with VLF codes one can approach capacity faster than in the fixed-blocklength case. Specifically, the \emph{channel dispersion}~\cite[Eq. (221)]{polyanskiy2010channel} turns out to be zero. The achievability bound used in~\cite{polyanskiy2011feedback} to prove this result is actually based on variable-length stop-feedback (VLSF) codes. In the VLSF setup, the feedback link is used by the receiver only to send a single bit indicating to stop the transmission of the current message. This setup is of interest from a practical point of view, because it encompasses hybrid automatic repetition request (ARQ) schemes. Note that VLSF codes are a  special case of VLF codes.

In this paper, we shall focus on the binary erasure channel (BEC) and seek nonasymptotic achievability and converse bounds, for both the VLF and the VLSF setups, which improve on the ones available in the literature. Note that the two-phase converse bounds~\cite[Thm. 1]{burnashev1976data},~\cite[Thm. 6]{polyanskiy2011feedback} require that all entries of the channel transition matrix of the DMC are strictly positive\footnote{This is required for~$C_1$ in~\eqref{expression_burnashev_reliability} to be finite.}---an assumption that does not hold for the BEC. Bounds on the maximum rate achievable over a BEC in the VLF setup are provided in~\cite[Thm. 7]{polyanskiy2011feedback}. The achievability bound is based on a simple scheme (also suggested in~\cite{ForneyExponentialBounds1968}) where each bit is repeated until it is received correctly. The converse bound can be seen as a variable-length analogue of Fano's inequality (see~\cite[Lemmas 1 and 2]{burnashev1976data}). This bound does not require $C_1$ to be finite.

The problem of constructing VLSF codes over a BEC reduces to problem of constructing rateless erasure codes. Thus, one can get achievability bounds on the maximum coding rate in the VLSF setup by analyzing the performance of family of rateless codes such as random linear fountain codes~\cite[Sec. 3]{mackay2005fountain}. The only converse bounds that are available for VLSF codes (stop feedback) hold also in the VLF setup (full feedback) to the best of the authors' knowledge. This is actually the case for both the maximum coding rate and the reliability function.

Our contributions in this paper are as follows:
\begin{itemize}[leftmargin=*]
\item We provide nonasymptotic converse and achievability bounds on the maximum coding rate of VLF codes over BECs, which improve upon the ones provided in~\cite[Thm. 7]{polyanskiy2011feedback}. Our converse bound relies on sequential hypothesis testing and is inspired by the meta-converse framework~\cite[Sec. III.E]{polyanskiy2010channel}; the achievability bound combines the simple repetition scheme used in~\cite[Thm. 7]{polyanskiy2011feedback} with variable-length Huffman coding. For the case of zero error probability, the achievability and converse bounds match.
\item For the VLSF setup, we provide nonasymptotic achievability bounds that improve on the one reported in~\cite[Thm. 3]{polyanskiy2011feedback}. The bounds are obtained by exploiting that, for a BEC, the decoder is able to identify the correct message whenever only a single codeword is compatible with the sequence of channel outputs received up to that point (a property noted previously in e.g.,~\cite{massay2007zeroerror}). The random coding argument used in one of the bounds utilizes linear codes. Hence, the resulting coding scheme can be seen as variable-length extension of random linear fountain codes.
\end{itemize}
\iflongversion\else Some proofs are omitted for space constraint; they can be found in~\cite{devassy2016nonasymptoticlong}. \fi
\subsubsection*{Notation} Uppercase curly letters denote sets. The~$n$-fold Cartesian product of a set~$\chinpspace$ is denoted by~$\chinpspace^n.$ Uppercase boldface letters denote random quantities and lightface letters denote deterministic quantities. The distribution of a random variable~$\chinp$ is denoted by~$P_\chinp.$ With~$\Bexpectation{\cdot}$ we denote expectation and with~$\Bcondexpectation{P}{\cdot}$ we stress that the expectation is with respect to the probability law~$P.$ The indicator function is denoted by~$\indicator{\cdot}$ and we use the symbol~$\binaryfinitefield$ to indicate the binary Galois field. With~$\chinp_m^n$ we denote the random vector with entries~$\parantheses{\chinp_m,\chinp_{m+1},\dots,\chinp_n}.$ Similarly,~$x_m^n$ stands for a deterministic vector with entries~$\parantheses{x_m,x_{m+1},\dots,x_n}.$ We shall often use the following function:
\begin{IEEEeqnarray}{rCl}
\lstar{x}&=&\floor{\log_2x}+2(1-2^{\floor{\log_2x}-\log_2x}),\qquad x\in\reals.\label{defnition_huffman_avg_length}
\end{IEEEeqnarray}
Here,~$\floor{\cdot}$ denotes the floor operator. Furthermore, we shall use~$\ceil{\cdot}$ to denote the ceil operator. We let~$\Bbernoullidist{p}$ denote a Bernoulli-distributed random variable with parameter~$p$ and~$\Bgeometricdist{p}$ a geometrically distributed random variable with parameter~$p.$ The binary entropy function~$\Bbinaryentropy{\cdot}$ is defined as follows:
\begin{IEEEeqnarray}{rCl}
\Bbinaryentropy{x}&=&-x\log_2 x - (1-x)\log_2(1-x),\iflongversion\qquad\else\quad\fi x\in\parantheses{0,1}.\IEEEeqnarraynumspace
\end{IEEEeqnarray}

\section{Definition}\label{section_system_model}
We consider a BEC with input alphabet~$\chinpspace=\curlybrac{0,1}$ and output alphabet~$\choutspace=\curlybrac{0,\erasure,1},$ where~$\erasure$ denotes an erasure. A VLF code for the BEC is defined as follows.
\begin{defn}\label{definition_vlf_codes}
	\textit{(\cite[Def. 1]{polyanskiy2011feedback})} An~$(\latency,\messagecount,\errorprob)$--VLF code, where~$\latency$ is a positive real,~$\messagecount$ is a positive integer, and~$\errorprob\in\closedclosedinterval{0,1},$ consists~of:
	\begin{enumerate}[leftmargin=*]
		\item A random variable~$\commonrand,$ defined on a set~$\commonrandspace$ with\footnote{The bound on the cardinality of~$\commonrandspace$ given in~\cite{polyanskiy2011feedback} (i.e.,~$\abs{\commonrandspace}\leq 3$) can be improved by using the Fenchel-Eggleston theorem~\cite[p. 35]{eggleston1958convexity} in place of Caratheodory's theorem.}~$\abs{\commonrandspace}\leq 2,$ whose realization is revealed to the encoder and the decoder before the start of transmission. The random variable~$\commonrand$ acts as common randomness and enables the use of randomized encoding and decoding strategies.
		\item A sequence of encoders~$f_n:\commonrandspace\times\messagespace\times\choutspace^{n-1}\functionto\chinpspace,n\geq 1$ that generate the channel inputs
			\begin{IEEEeqnarray}{rCl}
				\chinp_n &=& f_n\parantheses{\commonrand,\inpmessage,\chout_1^{n-1}}. \label{expression_encoder_bec_vlf}
			\end{IEEEeqnarray}
		Here,~$\inpmessage$ denotes the  message, which is uniformly distributed on~$\messagespace=\curlybrac{1,2,\dots,\messagecount}.$ Note that the channel input at time~$n$ depends on all previous channel outputs (full feedback).
		\item A sequence of decoders~$g_n:\commonrandspace\times\choutspace^n\functionto\messagespace$ that provide the estimate of~$\inpmessage$ at time~$n.$
		\item A nonnegative integer-valued random variable~$\stoppingtime,$ which is a stopping time of the filtration 
			\begin{IEEEeqnarray}{rCl}
				\mathcal{G}_n=\sigma\curlybrac{\commonrand,\chout_1^n} \label{expression_filtration_bec_vlf}
			\end{IEEEeqnarray}\vspace{-9pt}
			and satisfies
			\begin{IEEEeqnarray}{rCl}
				\Bexpectation{\stoppingtime} &\leq& \latency.\label{expression_stoppingtime_bec_vlf}
			\end{IEEEeqnarray}
		\item The final estimate~$\decoderoutput = g_\stoppingtime\parantheses{\commonrand,\chout_1^\stoppingtime}$ of~$\inpmessage,$ which satisfies the error-probability constraint 
			\begin{IEEEeqnarray}{rCl}
				\probof{\decoderoutput\neq\inpmessage}&\leq&\errorprob.\label{expression_prob_error_constraint_bec_vlf}
			\end{IEEEeqnarray}
	\end{enumerate}
\end{defn}
The rate~$\rate$ of an~$(\latency,\messagecount,\errorprob)$--VLF code is defined as
\begin{IEEEeqnarray}{rCl}
\rate &=& \frac{\log_2\messagecount}{\Bexpectation{\stoppingtime}}.\label{expression_rate_bec_vlf}
\end{IEEEeqnarray}
Furthermore, we define the minimum average blocklength of VLF codes with~$\messagecount$ codewords and error probability not exceeding~$\errorprob$ as follows:
\begin{IEEEeqnarray}{rCl}
\vlffundlatencylimit{\messagecount}{\errorprob}&=&\min\curlybrac{\latency\ :\ \exists(\latency,\messagecount,\errorprob)\text{--VLF code}}.\label{expression_min_avg_latency_vlf}
\end{IEEEeqnarray}
VLSF codes are a special case of VLF codes. The peculiarity of VLSF codes is that the sequence of encoders is not allowed to depend on the past channel outputs, i.e.,
\begin{IEEEeqnarray}{rCl}
f_n:\commonrandspace\times\messagespace\functionto\chinpspace,\ n\geq 1.\label{expression_encoder_bec_vlsf}
\end{IEEEeqnarray}
 In the VLSF case, the feedback link is used by the receiver only to inform the transmitter that the message has been decoded (stop/decision feedback).

\section{Existing Results for BEC}\label{section_existing_results}
In this section, we review the results available in literature on the minimum average blocklength~$\vlffundlatencylimit{\messagecount}{\errorprob}$ for the BEC. The following achievability bound is obtained by time-sharing between a scheme that drops the message to be transmitted without using the channel at all, and a scheme that repeats the channel input until it is received correctly.
\begin{thm}\label{theorem_achievability_BEC_yury_zero_prob}
\emph{(\cite[Thm. 7]{polyanskiy2011feedback})} For a BEC with erasure probability~$\becerasureprob,$ there exists an~$(\latency,\messagecount,\errorprob)$--VLF code with
\begin{IEEEeqnarray}{rCl}
\latency&\leq& \frac{\parantheses{1-\errorprob}\ceil{\log_2\messagecount}}{1-\becerasureprob}.\label{expression_achievability_BEC_yury_zero_prob}
\end{IEEEeqnarray}
\end{thm}
Next, we provide a converse bound.
\begin{thm}\label{theorem_converse_BEC_fano_zero_prob}
\emph{(\cite[Thm. 7]{polyanskiy2011feedback},\cite[Lemmas 1 and 2]{burnashev1976data})} For every~$(\latency,\messagecount,\errorprob)$--VLF code with~$0\leq\errorprob\leq 1-1/\messagecount$ operating over a BEC with erasure probability~$\becerasureprob,$ we have
\begin{IEEEeqnarray}{rCl}
\latency&\geq& \frac{(1-\errorprob)\log_2\messagecount-\Bbinaryentropy{\errorprob}}{1-\becerasureprob}.\label{expression_converse_BEC_fano_zero_prob}
\end{IEEEeqnarray}
\end{thm}
This converse bound can be obtained by constructing an appropriate martingale using the conditional entropy of the \emph{a posteriori} distribution of the message given the channel output. Note that the bounds~\eqref{expression_achievability_BEC_yury_zero_prob} and~\eqref{expression_converse_BEC_fano_zero_prob} coincide for~$\errorprob=0$ whenever the number of messages~$\messagecount$ is a power of~$2.$

\section{Novel Bounds for VLF Codes}\label{section_results_bec_vlf}
In this section, we present an achievability and a converse bound on~$\vlffundlatencylimit{\messagecount}{\errorprob}$ that improve upon the ones given in Theorems~\ref{theorem_achievability_BEC_yury_zero_prob} and~\ref{theorem_converse_BEC_fano_zero_prob}. The idea behind the achievability bound is to combine the scheme in Theorem~\ref{theorem_achievability_BEC_yury_zero_prob} with a Huffman code, whose purpose is to reduce the average blocklength when the number of messages is not a power of two. The converse bound relies on sequential hypothesis testing and is inspired by the meta-converse framework~\cite[Sec. III.E]{polyanskiy2010channel}. As we shall see, achievability and converse bounds are tight when~$\errorprob=0,$\emph{ for every} integer~$\messagecount.$ Our achievability bound is given in Theorem~\ref{theorem_achievability_bec_vlf} below.\ignorespaces
\begin{thm}\label{theorem_achievability_bec_vlf}
For a BEC with erasure probability~$\becerasureprob,$ there exists an~$(\latency,\messagecount,\errorprob)$--VLF code with 
\begin{IEEEeqnarray}{rCl}
\latency&\leq& \frac{\parantheses{1-\errorprob}\lstar{\messagecount}}{1-\becerasureprob}\label{expression_vlf_achievability}
\end{IEEEeqnarray}
where~$\lstar{\cdot}$ is defined in~\eqref{defnition_huffman_avg_length}.
\end{thm}
\begin{IEEEproof}
\iflongversion
See Appendix~\ref{appendix_proof_achievability_bec_vlf}.
\else
See~\cite[App. A]{devassy2016nonasymptoticlong}.
\fi
\end{IEEEproof}

The converse bound is based on binary sequential hypothesis testing~\cite{wald1945sequential}. Let\footnote{We use the same notation as in~\cite[Ch. 3]{tartakovsky2014sequential}.}~$\parantheses{\decisionrule,\stoppingtime}$ denote a generic binary sequential hypothesis test between two stochastic processes~$\processP$ and~$\processQ.$ Here,~$\stoppingtime$ is a stopping time and~$\decisionrule$ is a decision rule ($0$ indicates that~$\processP$ is chosen and~$1$ that~$\processQ$ is chosen). Let~$\Btaurulespace{\processP}{\processQ}$ denote the set of all possible binary sequential hypothesis tests. We are interested in the minimum average number of samples~$\Bprocessminavglatency{\testpowerP}{\testpowerQ}{\processP}{\processQ}$ required by a binary sequential hypothesis test to identify~$\processP$ and~$\processQ$ correctly with probability at least~$\testpowerP$ and~$\testpowerQ,$ respectively. Formally,
\begin{IEEEeqnarray}{rCl}
\Bprocessminavglatency{\testpowerP}{\testpowerQ}{\processP}{\processQ} &=& \min_{\substack{\parantheses{\decisionrule,\stoppingtime}\in\Btaurulespace{\processP}{\processQ},\\ P\squarebrac{\decisionrule=0}\geq\testpowerP,\\ Q\squarebrac{\decisionrule=1}\geq\testpowerQ}} \Bcondexpectation{P}{\stoppingtime}.\label{defnition_minlatency_given_testpower}
\end{IEEEeqnarray}
In Lemma~\ref{lemma_metaconverse_bec_vlf} below we establish a connection between~$\Bprocessminavglatency{\testpowerP}{\testpowerQ}{\processP}{\processQ}$ and the parameters of a given VLF code. For the sake of generality, the lemma is formulated for the case of arbitrary DMCs (this requires a suitable generalization of the definition of VLF codes provided in Definition~\ref{definition_vlf_codes} to arbitrary DMCs).
\begin{lem}\label{lemma_metaconverse_bec_vlf}
Consider an~$(\latency,\messagecount,\errorprob)$--VLF code for the DMC~$P_{\chout\given\chinp}.$ Let~$\errorprobQ$ denote the probability of error when this code is used over the DMC~$Q_{\chout\given\chinp}.$ Let~$\vlfprocessPygivenx$ and~$\vlfprocessQygivenx$ be the probability distribution of the process~$\commonrand,\curlybrac{\parantheses{\chinp_n,\chout_n}}_{n=1}^\infty$ under~$P_{\chout\given\chinp}$ and~$Q_{\chout\given\chinp},$ respectively. The distributions of the stochastic processes depend on the chosen~$(\latency,\messagecount,\errorprob)$--VLF code through its encoder according to~\eqref{expression_encoder_bec_vlf}. We consider binary sequential hypothesis testing between the two processes, under the assumption that the realization of~$\commonrand$ is known to the test before processing~$(\chinp_1,\chout_1).$ We have
\begin{IEEEeqnarray}{rCl}
\latency\geq \Bprocessminavglatency{1-\errorprob}{\errorprobQ}{\vlfprocessPygivenx}{\vlfprocessQygivenx}.\IEEEeqnarraynumspace\label{expression_metaconverse_bec_vlf}
\end{IEEEeqnarray}
\end{lem}
\begin{IEEEproof}\ignorespaces
See Appendix~\ref{appendix_proof_metaconverse_bec_vlf}.\ignorespaces
\end{IEEEproof}
The bound~\eqref{expression_metaconverse_bec_vlf} can be viewed as the variable-length analogue of the meta-converse theorem~\cite[Thm. 26]{polyanskiy2010channel}. The meta-converse theorem links the average error probabilities resulting by using the same fixed-blocklength code over two different channels by means of binary hypothesis testing. Similarly, Lemma~\ref{lemma_metaconverse_bec_vlf} relates the average error probabilities and the average blocklengths resulting by using a given VLF code over two different channels by means of binary sequential hypothesis testing.

To obtain a converse bound from~\eqref{expression_metaconverse_bec_vlf}, we take~$Q_{\chout\given\chinp}=Q_{\chout},$ with~$Q_{\chout}$ being the capacity-achieving output distribution of the BEC. Then, we solve the minimization in~\eqref{defnition_minlatency_given_testpower} by using the sequential probability ratio rest (SPRT)~\cite{wald1945sequential} (see\iflongversion Appendix~\ref{appendix_wald_sprt}\else~\cite[App. C]{devassy2016nonasymptoticlong} \fi for a short review). This yields the following bound.
\begin{thm}\label{theorem_converse_bec_vlf}
Every~$(\latency,\messagecount,\errorprob)$--VLF code operating over a BEC with erasure probability~$\becerasureprob$ satisfies
\begin{IEEEeqnarray}{rCl}
\latency &\geq& \frac{\parantheses{1-\errorprob}\lstar{\messagecount(1-\errorprob)}}{1-\becerasureprob}.\label{expression_converse_bec_vlf}
\end{IEEEeqnarray}
\end{thm}
\begin{IEEEproof}
\iflongversion
See Appendix~\ref{appendix_theorem_converse_bec_vlf}.
\else
See~\cite[App. D]{devassy2016nonasymptoticlong}.
\fi
\end{IEEEproof} 

We would like to emphasize that although~\eqref{expression_converse_bec_vlf} is tighter than the converse bound reported in~\cite[Thm. 6]{polyanskiy2011feedback}, a generalization of Theorem~\ref{theorem_converse_bec_vlf} to DMCs with finite~$C_1$ yields a converse bound that is in general looser than the ones reported in~\cite[Thm. 1]{burnashev1976data} and~\cite[Thm. 6]{polyanskiy2011feedback}. The peculiarity of the BEC is that the decoder is able to determine if its estimate is correct or not by assessing whether the estimated codeword  is the only codeword compatible with the erasure pattern. This implies that a two-phase scheme with a confirmation from the encoder is not required.

Note that the right-hand sides of~\eqref{expression_vlf_achievability} and~\eqref{expression_converse_bec_vlf} coincide when~$\errorprob=0.$ This fact is collected in the following corollary.\ignorespaces
\begin{cor}\label{corollary_vlf_min_latency_zero_error}
The minimum average blocklength~$\vlffundlatencylimit{\messagecount}{0}$ of an~$(\latency,\messagecount,0)$--VLF code over a BEC with erasure probability~$\becerasureprob$ is given by\ignorespaces
\begin{IEEEeqnarray}{rCl}
\vlffundlatencylimit{\messagecount}{0} &=& \frac{\lstar{\messagecount}}{1-\becerasureprob}\label{expression_vlf_min_latency}
\end{IEEEeqnarray}\ignorespaces
where~$\lstar{\cdot}$ is defined in~\eqref{defnition_huffman_avg_length}.
\end{cor}

\section{Novel Bounds for VLSF Codes}\label{section_vlsf_results}
We now focus on the VLSF setup and provide achievability bounds for the case~$\errorprob=0.$ Achievability bounds for arbitrary~$\errorprob$ can be obtained by allowing the receiver to send a stop signal at time zero with probability~$\errorprob$ (see~\cite[Sec. III.D]{polyanskiy2011feedback}). The corresponding achievability bounds can be readily obtained from the ones presented in this section by multiplying them by~$\parantheses{1-\errorprob}.$ 

As already mentioned, in the BEC case the decoder can assess the correctness of its message estimate by verifying whether the codeword corresponding to the chosen message is the only one that is compatible with the received sequence. It is therefore natural to consider a decoder whose stopping time is given by\ignorespaces
\begin{IEEEeqnarray}{rCl}
\stoppingtime &=& \inf \curlybrac{n\geq 1: \probof{\inpmessage = \Rdecoderoutput_n \given \chout_1^n = \Rchout_1^n} = 1 } \label{expression_bec_vlsf_optimal_stopping_time}
\end{IEEEeqnarray}\ignorespaces
where~$\Rdecoderoutput_n$ denotes the message estimate at the decoder after~$n$ channel uses:
\begin{IEEEeqnarray}{rCl}
\Rdecoderoutput_n &=& \argmax_{w\in\messagespace} \probof{\inpmessage=w\given \chout_1^n = \Rchout_1^n}.\label{expression_bec_vlsf_optimal_decoder}
\end{IEEEeqnarray}
The decoding rule~\eqref{expression_bec_vlsf_optimal_stopping_time}--\eqref{expression_bec_vlsf_optimal_decoder} combined with random coding (independent and identically distributed (\iid)~$\Bbernoullidist{0.5}$ ensemble) yields the following achievability bound.\ignorespaces
\begin{thm}\label{theorem_bec_achievability_iid_no_union_bound}
For a BEC with erasure probability~$\becerasureprob,$ there exists an~$(\latency,\messagecount,0)$--VLSF code with 
\begin{IEEEeqnarray}{rCl}
\latency &\leq& \frac{1}{1-\becerasureprob}\parantheses{1-\sum_{i=1}^{\messagecount-1}\binom{\messagecount-1}{i}\frac{\parantheses{-1}^i}{2^i-1}}.\label{expression_bec_achievability_iid_no_union_bound}
\end{IEEEeqnarray}
\end{thm}
\begin{IEEEproof}\ignorespaces
See Appendix~\ref{appendix_proof_bec_achievability_iid_no_union_bound}.\ignorespaces
\end{IEEEproof}
The achievability bound~\eqref{expression_bec_achievability_iid_no_union_bound} suffers from two pitfalls: (i) the bound is loose when~$\messagecount$ is small because the random coding ensemble contains few codebooks with abnormally large average blocklengths; (ii) since the bound requires the computation of differences of binomial coefficients, it becomes difficult to compute when~$\messagecount$ is larger than~$10^4.$ Next, we present a different achievability bound that addresses these two shortcomings. To tighten~\eqref{expression_bec_achievability_iid_no_union_bound} for small~$\messagecount$ we use an expurgation technique similar to the ones utilized by Shannon, Gallager, and Berlekamp~\cite[p. 529]{ShannonGallagher1967522}. Specifically, we view each codebook as a random matrix with~$\messagecount$ rows and an infinite number of columns and we assign the following probability distribution on the VLSF code ensemble: each column is drawn uniformly and independently from the set of binary vectors with~$\ceil{\messagecount/2}$ zeros. Furthermore, to obtain an expression that is computable for arbitrary values of~$\messagecount,$ we upper-bound the average blocklength of the expurgated ensemble using the union bound. The achievability bound thus obtained is given in the following theorem.\ignorespaces
\begin{thm}\label{theorem_bec_achiveability_vlsf_zero_error}
For a BEC with erasure probability~$\becerasureprob,$ there exists an~$(\latency,\messagecount,0)$--VLSF code with 
\begin{IEEEeqnarray}{rCl}
\latency&\leq& \frac{1}{1-\becerasureprob}\parantheses{\floor{m}+1+\frac{\invprobofunequalbits^{m-\floor{m}}}{\invprobofunequalbits-1}} \label{expression_bec_achiveability_vlsf_zero_error}
\end{IEEEeqnarray}
where~$m=\log_\invprobofunequalbits\parantheses{\messagecount-1}$ and~$\invprobofunequalbits$ is related to the number of messages~$\messagecount$ as follows:
\begin{IEEEeqnarray}{rCl}
\invprobofunequalbits &=& 2+\frac{1}{\ceil{\messagecount/2}-1}.\label{expression_inv_prob}
\end{IEEEeqnarray}
\end{thm}
\begin{IEEEproof}
\iflongversion
See Appendix~\ref{appendix_proof_bec_achiveability_vlsf_zero_error}.
\else
See~\cite[App. G]{devassy2016nonasymptoticlong}.
\fi
\end{IEEEproof}
The parameter~$\invprobofunequalbits$ in~\eqref{expression_inv_prob} is the reciprocal of the probability that the first two bits in a random vector that is uniformly distributed over the set of vectors in~$\binaryfinitefield^{\messagecount}$ with~$\ceil{\messagecount/2}$ zeros, are equal.

For the case when the number of messages~$\messagecount$ is a power of~$2,$ one can obtain an achievability bound that is tighter than~\eqref{expression_bec_achiveability_vlsf_zero_error}, that does not require the union bound, and that is easily computable. The bound relies on a linear codebook ensemble in which the columns of the random generator matrix are distributed uniformly over the set of all nonzero vectors from~$\binaryfinitefield^{\log_2\mathopen{}\mathclose\messagecount}.$ Specifically, consider a received vector of length~$n$ and let~$\indexset_n$ be the set containing the indices of unerased symbols in the received vector. We are interested in the dimension of the subspace spanned by the columns of the generator matrix with index in~$\indexset_n.$ This dimension evolves as a Markov chain with a single absorbing state (the state that corresponds to maximum dimension~$\log_2\mathopen{}\mathclose\messagecount$). it follows that the average blocklength (averaged over the ensemble) coincides with the expected absorption time of the Markov chain, which follows a discrete phase-type distribution~\cite[Ch. 2]{neuts1981matrix}. This achievability scheme can be seen as a variable-length analogue of random linear fountain codes~\cite[Sec. 3]{mackay2005fountain}. The performance of the achievability scheme just described is characterized in the following theorem.\ignorespaces
\begin{thm}\label{theorem_bec_achievability_expurgated_linear}
For every integer~$\messagebitcount\geq1,$ there exists an~$(\latency,2^\messagebitcount,0)$--VLSF code for a BEC with erasure probability~$\becerasureprob$ with 
\begin{IEEEeqnarray}{rCl}
\latency &\leq& \frac{1}{1-\becerasureprob}\parantheses{\messagebitcount+\sum_{i=1}^{\messagebitcount-1}\frac{2^{i}-1}{2^k-2^{i}}}.\label{expression_bec_achievability_expurgated_linear}
\end{IEEEeqnarray}
\end{thm}
\begin{IEEEproof}
\iflongversion
See Appendix~\ref{appendix_proof_bec_achievability_expurgated_linear}.
\else
See~\cite[App. H]{devassy2016nonasymptoticlong}.
\fi
\end{IEEEproof}

\begin{figure}
\centerline{\includegraphics{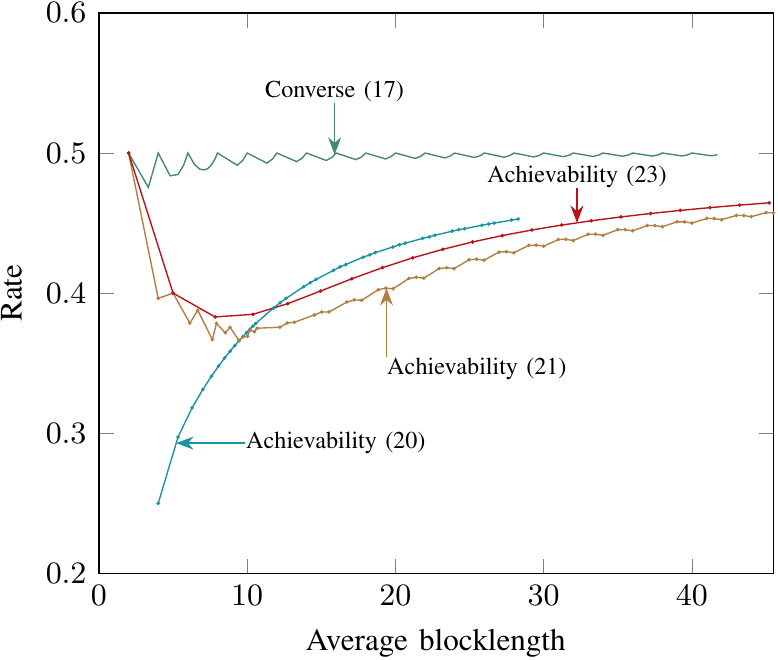}}
\caption{Achievability and converse bounds for zero error VLSF codes over BEC with~$\becerasureprob=0.5.$ The converse bound is the one given in~\eqref{expression_vlf_min_latency} for VLF codes.}\label{fig_vlsf_bec}\vspace{-15pt}\ignorespaces
\end{figure}

The achievability bounds~\eqref{expression_bec_achievability_iid_no_union_bound},~\eqref{expression_bec_achiveability_vlsf_zero_error}, and~\eqref{expression_bec_achievability_expurgated_linear} are plotted in Fig.~\ref{fig_vlsf_bec}. \ignorespaces
\ignorespaces
\iftwocol 
\begin{figure*}[!ht]\ignorespaces
\begin{IEEEeqnarray}{rCl}
\IEEEeqnarraymulticol{3}{l}{\probof{\errorevent=0\given \chinp_1^n=\Rchinp_1^n,\chout_1^n=\Rchout_1^n,\commonrand=\Rcommonrand,\stoppingtime=n} }\nonumber\\
&=&  \sum_{w\in\messagespace}\probof{\inpmessage=w\given \chinp_1^n=\Rchinp_1^n,\chout_1^n=\Rchout_1^n,\commonrand=\Rcommonrand}\probof{\decoderoutput=w\given \chout_1^n=\Rchout_1^n,\commonrand=\Rcommonrand,\stoppingtime=n}.\IEEEeqnarraynumspace\label{expression_errorevent_conditioned}
\end{IEEEeqnarray}\ignorespaces
\hrulefill\ignorespaces\vspace{-10pt}
\end{figure*}\ignorespaces
\fi\ignorespaces
As expected, the bound~\eqref{expression_bec_achiveability_vlsf_zero_error} is tighter than~\eqref{expression_bec_achievability_iid_no_union_bound} at small average blocklengths (because of expurgation) and looser at large average blocklengths (because of union bound). The achievability bound~\eqref{expression_bec_achievability_expurgated_linear} is tighter than~\eqref{expression_bec_achiveability_vlsf_zero_error} for all blocklengths and looser than~\eqref{expression_bec_achievability_iid_no_union_bound} for large average blocklengths. When~$\messagecount=2,$ the achievability bounds~\eqref{expression_bec_achiveability_vlsf_zero_error} and~\eqref{expression_bec_achievability_expurgated_linear} coincide with the converse bound for VLF codes given in~\eqref{expression_vlf_min_latency}. This holds because the scheme that achieves~\eqref{expression_vlf_min_latency} (repeat each bit until it is received correctly) can be implemented with stop feedback when\footnote{Recall that in the VLSF setup the decoder is allowed to send only one stop signal per message.}~$\messagecount=2.$ As~$\messagecount$ increases, the gap between the VLSF achievability bounds and the VLF converse bound increases and gets as large as~$23\%$ when~$\messagecount=8,$ before vanishing asymptotically as~$\messagecount\tendsto\infty.$ It remains to be seen whether this gap is fundamental.\ignorespaces
%

\begin{appendices}
\iflongversion
\section{Proof of Theorem~\ref{theorem_achievability_bec_vlf}}\label{appendix_proof_achievability_bec_vlf}
We use time sharing between a scheme that drops the message to be transmitted without using the channel at all, and a zero-error VLF code constructed as follows: we first generate a prefix-free Huffman code~\cite{huffman1952method} for the~$\messagecount$ equiprobable messages. To send a given message, we repeat each bit of the corresponding Huffman codeword until it is received correctly (note that this requires full feedback at the transmitter). The average blocklength of the resulting VLF code can be analyzed as follows. Let~$\huffmanavglength{\messagecount}$ denote the average code length of the Huffman code for the~$\messagecount$ equiprobable messages. The average blocklength of the VLF code resulting from our construction is given by 
\begin{IEEEeqnarray}{rCl}
\latency&=&\frac{\huffmanavglength{\messagecount}}{1-\becerasureprob}.\label{expression_huffman_avg_latency}
\end{IEEEeqnarray}
Note now that the length of the Huffman codeword assigned to each message is either~$\floor{\log_2\messagecount}$ or~$\floor{\log_2\messagecount}+1$ (see~\cite[p. 598]{ahlswede2006general}). Specifically, the Huffman code assigns codewords of length~$\floor{\log_2\messagecount}$ to~$2^{\floor{\log_2\messagecount}+1}-\messagecount$ messages, and codewords of length~$\floor{\log_2\messagecount}+1$ to the remaining messages. This allows us to conclude that
\begin{IEEEeqnarray}{rCl}
\huffmanavglength{\messagecount}&=&\lstar{\messagecount}\label{expression_huffman_avg_codelength}
\end{IEEEeqnarray}
where~$\lstar{\cdot}$ is defined in~\eqref{defnition_huffman_avg_length}. We obtain~\eqref{expression_vlf_achievability} by allowing the transmitter to drop each codeword with probability~$\errorprob$ (see~\cite[Sec. III.D]{polyanskiy2011feedback}).
\fi
\section{Proof of Lemma~\ref{lemma_metaconverse_bec_vlf}}\label{appendix_proof_metaconverse_bec_vlf}
\iftwocol\ignorespaces
Consider the random variable~$\errorevent=\indicator{\inpmessage\neq\decoderoutput}.$ The conditional distribution of~$\errorevent$ given $\chinp_1^n=\Rchinp_1^n,\chout_1^n=\Rchout_1^n,\commonrand=\Rcommonrand,\stoppingtime=n,$ which is given in~\eqref{expression_errorevent_conditioned}, does not depend on whether the underlying channel is~$P_{\chout\given\chinp}$ or~$Q_{\chout\given\chinp}.$ 
Indeed in~\eqref{expression_errorevent_conditioned}, the first factor  depends only on the encoder and second factor depends only on the decoder.
\else 
Consider the random variable~$\errorevent=\indicator{\inpmessage\neq\decoderoutput}.$ The conditional distribution of~$\errorevent$ given $\chinp_1^n=\Rchinp_1^n,\chout_1^n=\Rchout_1^n,\commonrand=\Rcommonrand,\stoppingtime=n$ does not depend on whether the underlying channel is~$P_{\chout\given\chinp}$ or~$Q_{\chout\given\chinp}.$ Indeed,
\begin{IEEEeqnarray}{rCl}
\IEEEeqnarraymulticol{3}{l}{\probof{\errorevent=0\given \chinp_1^n=\Rchinp_1^n,\chout_1^n=\Rchout_1^n,\commonrand=\Rcommonrand,\stoppingtime=n} }\nonumber\\
&=& \sum_{w\in\messagespace}\probof{\inpmessage=w\given \chinp_1^n=\Rchinp_1^n,\chout_1^n=\Rchout_1^n,\commonrand=\Rcommonrand}\probof{\decoderoutput=w\given \chout_1^n=\Rchout_1^n,\commonrand=\Rcommonrand,\stoppingtime=n}\IEEEeqnarraynumspace\label{expression_errorevent_conditioned}
\end{IEEEeqnarray}
where the first factor depends only on the encoder and second factor depends only on the decoder. 
\fi
Using the stopping time~$\stoppingtime$ associated to the given code and the family of probability kernels defined by the conditional distribution~\eqref{expression_errorevent_conditioned} we construct a binary sequential hypothesis test~$\parantheses{\decisionrule,\stoppingtime}$. By definition, we have that under~$P_{\chout\given\chinp},$
{\setlength{\belowdisplayskip}{0pt} \setlength{\belowdisplayshortskip}{0pt}
\begin{IEEEeqnarray}{rCl}
\probof{\decisionrule=0} &=& 1-\errorprob
\end{IEEEeqnarray} and under~$Q_{\chout\given\chinp}$
\begin{IEEEeqnarray}{rCl}
\probof{\decisionrule=0}&=& 1-\errorprobQ.
\end{IEEEeqnarray} Thus,
\begin{IEEEeqnarray}{rCl}
\latency &\geq&\Bcondexpectation{\vlfprocessPygivenx}{\stoppingtime}\\
&\geq&\Bprocessminavglatency{1-\errorprob}{\errorprobQ}{\vlfprocessPygivenx}{\vlfprocessQygivenx}.\IEEEeqnarraynumspace
\end{IEEEeqnarray}}\ignorespaces\vspace{-10pt}
\iflongversion

\section{Sequential Probability Ratio Test (SPRT)}\label{appendix_wald_sprt}\ignorespaces
In this appendix, we provide a brief overview of the sequential probability ratio test (SPRT)~\cite{wald1945sequential} and discuss its optimality. Let~$\parantheses{\decisionrule,\stoppingtime}$ denote a generic binary sequential hypothesis test between two stationary memoryless stochastic processes  with marginal distribution~$P_{\chinp}$ and~$Q_\chinp.$ Here,~$\stoppingtime$ is a stopping time---a random variable denoting the number of samples taken before making a decision---and~$\decisionrule$ is a decision rule ($0$ indicating that~$P_{\chinp}$ is chosen and~$1$ that~$Q_{\chinp}$ is chosen). With~$\Bcondexpectation{P_\chinp}{\stoppingtime}$ and~$\Bcondexpectation{Q_\chinp}{\stoppingtime}$ we denote the average number of samples required under the hypothesis~$P_{\chinp}$ and~$Q_\chinp,$ respectively. The probability of correct decision under hypothesis~$P_{\chinp}$ and~$Q_\chinp$  are denoted by~$P_\chinp\parantheses{\decisionrule=0}$ and~$Q_\chinp\parantheses{\decisionrule=1},$ respectively.

\begin{defn}\label{defnition_wald_sprt}
Let the log-likelihood ratio (LLR) after~$n$ samples be recursively defined as follows:
\begin{IEEEeqnarray}{rCl}
S_0&=&0\\
S_n&=&S_{n-1}+\Blog{\frac{\infinitesimal P_{\chinp}}{\infinitesimal Q_{\chinp}}(\Rchinp_n)},\qquad n\geq 1
\end{IEEEeqnarray}
where~$\Rchinp_n$ is the~$n$th sample and~$\frac{\infinitesimal P_{\chinp}}{\infinitesimal Q_{\chinp}}(\cdot)$ denotes the Radon-Nikodym derivative. Let~$A_Q$ and~$A_P$ be two nonnegative scalars. The SPRT with stopping bounds~$A_Q$ and~$A_P$ is defined as follows: 
\begin{IEEEeqnarray}{rCl}
\stoppingtime &=& \min\curlybrac{n\geq 0, S_n \notin \parantheses{-A_Q,A_P} } \label{expression_wald_sprt_stopping_rule}\\
\decisionrule &=& \left\{ \,
\begin{IEEEeqnarraybox}[][c]{l?l}
\IEEEstrut
1, &  S_\stoppingtime\leq -A_Q \\
0, & S_\stoppingtime\geq A_P.
\IEEEstrut 
\end{IEEEeqnarraybox}\right.\label{expression_wald_sprt_decision_rule}
\end{IEEEeqnarray}
\end{defn}

We denote the SPRT defined through~\eqref{expression_wald_sprt_stopping_rule} and~\eqref{expression_wald_sprt_decision_rule} by~$S(A_Q,A_P).$ Note that SPRT allows for the possibility that~$\stoppingtime=0$ (the test stops before processing the first sample). Specifically, the tests~$S(0,A_P)$ and~$S(A_Q,0)$ will stop at~$\stoppingtime=0$ (recall that~$S_0=0$) and declare the hypothesis to be~$Q_{\chinp}$ and~$P_\chinp,$ respectively. It will turn out convenient to denote with~$S(A_Q-,A_P)$ the test with stopping time given by
\begin{IEEEeqnarray}{rCl}
\stoppingtime &=& \min\curlybrac{n\geq 0, S_n \notin \closedopeninterval{-A_Q,A_P} }
\end{IEEEeqnarray}
and decision rule~\eqref{expression_wald_sprt_decision_rule}. Similarly, we use~$S(A_Q,A_P+)$ to denote the test with stopping time
\begin{IEEEeqnarray}{rCl}
\stoppingtime &=& \min\curlybrac{n\geq 0, S_n \notin \openclosedinterval{-A_Q,A_P} }
\end{IEEEeqnarray}
and decision rule~\eqref{expression_wald_sprt_decision_rule}. Finally, the test with stopping time
\begin{IEEEeqnarray}{rCl}
\stoppingtime &=& \min\curlybrac{n\geq 0, S_n \notin \closedclosedinterval{-A_Q,A_P} }
\end{IEEEeqnarray}
and decision rule~\eqref{expression_wald_sprt_decision_rule} is denoted by~$S(A_Q-,A_P+).$ 

A randomization between a finite collection of binary sequential hypothesis tests refers to a testing procedure where a test is randomly selected from the collection according to a given probability law. Such a testing procedure is also referred to as randomized test. Next, we provide an extension of SPRT that allows for randomization.
\begin{defn}\label{defnition_extended_sprt}
The \emph{extended} SPRT~\cite{Burkholder1963ptimumProperties} with stopping bounds~$A_Q,A_P$ is the test that chooses~$\decisionrule=1$ and~$\decisionrule=0$ when~$S_n<-A_Q$ and~$S_n>A_P,$ respectively, and requests the next sample when~$-A_Q<S_n<A_P.$ When~$S_n=-A_Q,$ a possibly randomized rule is adopted to decide whether to set~$\decisionrule=1$ or to request the next sample. Similarly, when~$S_n=A_P,$ a possibly randomized rule is adopted to decide whether to set~$\decisionrule=0$ or to request the next sample.
\end{defn}

As noted in~\cite[Rem. 2.1]{Burkholder1963ptimumProperties}, every randomized test obtained by randomizing between~$S(A_Q,A_P),$ $S(A_Q-,A_P),$ $S(A_Q,A_P+),$ and~$S(A_Q-,A_P+)$ is an extended SPRT with stopping bounds~$A_Q,A_P$. Next, we shall define the following \emph{optimum property}.
\begin{defn}\label{defnition_optimum_property}
A binary sequential hypothesis test~$(\decisionrule^*,\stoppingtime^*)$ is said to have optimum property~(OP) if~$\Bcondexpectation{P_\chinp}{\stoppingtime^*}$ and~$\Bcondexpectation{Q_\chinp}{\stoppingtime^*}$ are finite, and for every other test~$(\decisionrule,\stoppingtime)$ with finite~$\Bcondexpectation{P_\chinp}{\stoppingtime}$ and~$\Bcondexpectation{Q_\chinp}{\stoppingtime}$ the conditions
\begin{IEEEeqnarray}{rCl}
P_\chinp\parantheses{\decisionrule=0} &\geq&P_\chinp\parantheses{\decisionrule^*=0}\\
Q_\chinp\parantheses{\decisionrule=1} &\geq&Q_\chinp\parantheses{\decisionrule^*=1}
\end{IEEEeqnarray}
imply that
\begin{IEEEeqnarray}{rCl}
\Bcondexpectation{P_\chinp}{\stoppingtime^*}&\leq&\Bcondexpectation{P_\chinp}{\stoppingtime}\\
\Bcondexpectation{Q_\chinp}{\stoppingtime^*}&\leq&\Bcondexpectation{Q_\chinp}{\stoppingtime}.
\end{IEEEeqnarray}

\end{defn}

In~\cite{Wald1948OptimumCharacter} it is proven that every SPRT has OP. This result was later extended in~\cite[Cor. 2.1]{Burkholder1963ptimumProperties} to prove that every extended SPRT has OP as well. Since for every~$\testpowerP$ and~$\testpowerQ$ in~\eqref{defnition_minlatency_given_testpower} we can find an extended SPRT that satisfies~$P\squarebrac{\decisionrule=0}=\testpowerP$ and~$Q\squarebrac{\decisionrule=1}=\testpowerQ,$  we conclude that extended SPRT minimizes~\eqref{defnition_minlatency_given_testpower}. Note that randomization is in general required to guarantee that the test achieves~$P\squarebrac{\decisionrule=0}=\testpowerP$ and~$Q\squarebrac{\decisionrule=1}=\testpowerQ$ for every arbitrary pair of probabilities~$\testpowerP$ and~$\testpowerQ.$

\fi\iflongversion
\section{Proof of Theorem~\ref{theorem_converse_bec_vlf}}\label{appendix_theorem_converse_bec_vlf}

We use Lemma~\ref{lemma_metaconverse_bec_vlf} with~$Q_{\chout\given\chinp}=Q_{\chout},$ where~$Q_{\chout}$ is the capacity-achieving output distribution of the BEC, i.e.,
\begin{IEEEeqnarray}{rCl}
Q_{\chout}(0) &=& Q_{\chout}(1) = \frac{1-\becerasureprob}{2}\\
Q_{\chout}(\erasure) &=& \becerasureprob.
\end{IEEEeqnarray}
The error probability~$\errorprobQ$ of a given~$(\latency,\messagecount,\errorprob)$--VLF code over the channel~$Q_{\chout}$ can be evaluated as follows:
\begin{IEEEeqnarray}{rCl}
\IEEEeqnarraymulticol{3}{l}{1-\errorprobQ}\nonumber\\
&=&\sum_{\Rcommonrand\in\commonrandspace}\sum_{n=0}^\infty\sum_{\substack{\Rchinp_1^n\in\chinpspace^n \\ \Rchout_1^n\in\choutspace^n}}\sum_{w\in\messagespace}\frac{P_{\commonrand}\squarebrac{\Rcommonrand}\prod_{k=1}^n P_{\chinp_k \given \chout_1^{k-1},\inpmessage,\commonrand}\squarebrac{\Rchinp_k\given \Rchout_1^{k-1},w,\Rcommonrand} Q_{\chout_1^n}\squarebrac{\Rchout_1^n} P_{\decoderoutput,\stoppingtime \given \chout_1^n,\commonrand}\squarebrac{w,n\given \Rchout_1^n,\Rcommonrand}}{\messagecount}\nonumber\\
\label{expression_full_eps_q}
\\
&=&\sum_{\Rcommonrand\in\commonrandspace}\sum_{n=0}^\infty\sum_{\Rchout_1^n\in\choutspace^n}\sum_{w\in\messagespace}\frac{P_{\commonrand}\squarebrac{\Rcommonrand}Q_{\chout_1^n}\squarebrac{\Rchout_1^n}P_{\decoderoutput,\stoppingtime \given \chout_1^n,\commonrand}\squarebrac{w,n\given \Rchout_1^n,\Rcommonrand}}{\messagecount}\label{expression_xgone_eps_q}\\
&=&\sum_{\Rcommonrand\in\commonrandspace}\sum_{n=0}^\infty\frac{P_{\commonrand}\squarebrac{\Rcommonrand}P_{\stoppingtime\given \commonrand}\squarebrac{n\given \Rcommonrand}}{\messagecount}= \frac{1}{\messagecount}.\label{expression_full_eps_q_last_step}
\end{IEEEeqnarray}
To obtain~\eqref{expression_xgone_eps_q}, we used that
\begin{IEEEeqnarray}{rCl}
\sum_{\Rchinp_1^n\in\chinpspace^n}\prod_{k=1}^n f_k(x_k) &=& \prod_{k=1}^n\sum_{\Rchinp_k\in\chinpspace} f_k(x_k)
\end{IEEEeqnarray}
where~$f_k(\cdot)$ are arbitrary functions. Thus, we have 
\begin{IEEEeqnarray}{rCl}
\errorprobQ &=& 1-1/\messagecount.
\end{IEEEeqnarray}

We now proceed to solve the minimization in~\eqref{defnition_minlatency_given_testpower} for the case~$\testpowerP=1-\errorprob,\testpowerQ=1-1/\messagecount,P=\vlfprocessPygivenx,Q=\vlfprocessQygivenx.$ Here,~$\vlfprocessQygivenx$ denotes the distribution of the stochastic process~$\commonrand,\curlybrac{\parantheses{\chinp_n,\chout_n}}_{n=1}^{\infty}$ under channel~$Q_\chout.$ Specifically,
\begin{IEEEeqnarray}{rCl}
Q_{\commonrand,\chinp_1^n,\chout_1^n}\squarebrac{\Rcommonrand,\Rchinp_1^n,\Rchout_1^n}&=& \sum_{\Rinpmessage\in\messagespace} P_\commonrand\squarebrac{\Rcommonrand}P_{\inpmessage\given\commonrand}\squarebrac{\Rinpmessage\given\Rcommonrand}\prod_{k=1}^{n}P_{\chinp_k\given\inpmessage,\commonrand,\chout_1^{k-1}}\squarebrac{\Rchinp_k\given\Rinpmessage,\Rcommonrand,\Rchout_1^{k-1}}Q_{\chout}\squarebrac{\Rchout_k}.\IEEEeqnarraynumspace
\end{IEEEeqnarray}
 We shall show that the binary sequential hypothesis test that achieves the minimum in~\eqref{defnition_minlatency_given_testpower} is the extended SPRT reviewed in Appendix~\ref{appendix_wald_sprt}. To do so, we need to show that the LLR process (see Def.~\ref{defnition_wald_sprt} in Appendix~\ref{appendix_wald_sprt})
\begin{IEEEeqnarray}{c}
\curlybrac{\llr_n=\log{\frac{P_{\commonrand,\chinp_1^n,\chout_1^n}}{Q_{\commonrand,\chinp_1^n,\chout_1^n}}}}_{n=0}^{\infty}
\end{IEEEeqnarray}
is a process with~\iid increments~\cite[p. 157]{tartakovsky2014sequential}. Indeed, consider the following quantity:
\begin{IEEEeqnarray}{rCl}
\conditionalllr_n &=& \log\frac{P_{\chout_n\given\chinp_n}\squarebrac{\Rchout_n\given\Rchinp_n}}{Q_{\chout_n}\squarebrac{\Rchout_n}}.
\end{IEEEeqnarray}
One can verify that~$\llr_n$ can be expressed as follows:
\begin{IEEEeqnarray}{rCl}
\bS_0&=&0\\
\bS_n&=&\bS_{n-1}+\conditionalllr_n,\qquad n\geq 1.
\end{IEEEeqnarray}
Note now that under~$\vlfprocessPygivenx$ the distribution of $\conditionalllr_n$ is
\begin{IEEEeqnarray}{lCl}
\probof{\conditionalllr_n =\log 2}&=&1-\becerasureprob\label{expression_condionalllr_pmf_p_log2}\\
\probof{\conditionalllr_n =0}&=&\becerasureprob.\label{expression_condionalllr_pmf_p_0}
\end{IEEEeqnarray}
Furthermore, under~$\vlfprocessQygivenx$ the distribution of $\conditionalllr_n$ is
\begin{IEEEeqnarray}{lCl}
\probof{\conditionalllr_n =\log 2}&=&\frac{1-\becerasureprob}{2}\label{expression_condionalllr_pmf_q_log2}\\
\probof{\conditionalllr_n =0}&=&\becerasureprob\label{expression_condionalllr_pmf_q_0}\\
\probof{\conditionalllr_n =-\infty}&=&\frac{1-\becerasureprob}{2}.\label{expression_condionalllr_pmf_q_minusinfinity}
\end{IEEEeqnarray}
Moreover, the stochastic process~$\curlybrac{\conditionalllr_n}_{n=1}^{\infty}$ is~\iid under both distributions. This allows us to conclude that extended SPRT achieves the minimum in~\eqref{defnition_minlatency_given_testpower} (see~\cite[Sec. 3.2.3]{tartakovsky2014sequential}). Such a test will be  a randomization between the following tests (see Appendix~\ref{appendix_wald_sprt} for a clarification on the notation used here):
\begin{IEEEeqnarray}{rCl}
S(0,m\log 2),S(0,m\log 2+),S(0-,m\log 2),S(0-,m\log 2+).
\end{IEEEeqnarray}
Here,~$m$ is a positive integer. Note that the tests~$S(0,m\log 2)$ and~$S(0,m\log 2+)$ stop at~$\stoppingtime=0$ and choose~$\vlfprocessQygivenx$. Furthermore, the test~$S(0-,m\log 2+)$ coincides with the test~$S(0-,(m+1)\log 2).$ 

The probability of correct decision~$\testpowerQ$ under hypothesis~$\vlfprocessQygivenx$ for the test~$S(0-,m\log 2)$ is given by
\begin{IEEEeqnarray}{rCl}
\testpowerQ&=&1-\parantheses{\frac{1-\becerasureprob}{2}}^m \sum_{n=0}^{\infty}  \binom{n+m-1}{m-1} \becerasureprob^{n} \label{proof_step_bionmial_series}\\
&=& 1-2^{-m}. \label{proof_step_after_binomial_series}
\end{IEEEeqnarray}
In~\eqref{proof_step_bionmial_series}, we used that the probability of error under~$\vlfprocessQygivenx$ is the probability that a sequence of~\iid ternary random variables distributed according to~\eqref{expression_condionalllr_pmf_q_log2}-\eqref{expression_condionalllr_pmf_q_minusinfinity} has~$m$ entries equal to~$\log 2$ (one of them being in the last position) and all remaining entries equal to~$0.$ We obtain~\eqref{proof_step_after_binomial_series} by using that the sum in~\eqref{proof_step_bionmial_series} is the binomial series expansion of~$1/(1-\becerasureprob)^{m}.$ With similar steps, one can prove that the average number of samples under~$\vlfprocessPygivenx$ that are required for the test~$S(0-,m\log 2)$ to stop is 
\begin{IEEEeqnarray}{rCl}
\Bcondexpectation{\vlfprocessPygivenx}{\stoppingtime} &=& \frac{m}{1-\becerasureprob}.
\end{IEEEeqnarray}
Finally, we obtain~\eqref{expression_converse_bec_vlf} by imposing that~$\testpowerP=1-\errorprob$ and~$\testpowerQ=1-1/\messagecount,$ and by solving for the integer~$m$ and the randomization probabilities.

\fi\iflongversion
\section{An Auxiliary Result}\label{appendix_auxiliary_results}
In this appendix, we provide a lemma that is used in the proof of Theorems~\ref{theorem_bec_achievability_iid_no_union_bound}--\ref{theorem_bec_achievability_expurgated_linear}. We consider the evaluation of the average blocklength, averaged over a given ensemble of VLSF codes that operate over a BEC with erasure probability~$\becerasureprob>0.$ The lemma allows us to relate this average blocklength to the one corresponding to the case~$\becerasureprob=0.$
\begin{lem}\label{lemma_reduce_to_zero_erasure}
Consider a BEC with erasure probability~$\becerasureprob>0$ and the ensemble of~$(\latency,\messagecount,0)$--VLSF codes constructed as follows: the stopping time and the decoder are defined as in~\eqref{expression_bec_vlsf_optimal_stopping_time} and~\eqref{expression_bec_vlsf_optimal_decoder}, respectively; the codebook of each code---a matrix with~$\messagecount$ rows and infinitely many columns---has columns independently generated from a given~$\messagecount$ dimensional probability distribution. Let~$\stoppingtime_0$ be the stopping time when~$\becerasureprob=0.$ Then
\begin{IEEEeqnarray}{rCl}
\Bexpectation{\stoppingtime} &=& \frac{\Bexpectation{\stoppingtime_0}}{1-\becerasureprob}\label{expression_reduce_to_zero_erasure}
\end{IEEEeqnarray}
where the expectation is  over both channel and code ensemble.
\end{lem}
\begin{IEEEproof}
Let~$\stoppingtime$ be the random variable corresponding to the length of a sequence of output symbols for which the decoder stops. Let~$\stoppingtime_0$ be the number of unerased symbols in the sequence. The expected value of~$\stoppingtime_0$ averaged with respect to both code ensemble and channel law coincide with the average stopping time when~$\becerasureprob=0.$ Let now~$\unerasedposition_1,\unerasedposition_2,\dots,\unerasedposition_{\stoppingtime_0}$ be the position of the unerased bits in the sequence of output symbols. Let~$\erasurecount_1=\unerasedposition_1$ and $\erasurecount_n=\unerasedposition_n-\unerasedposition_{n-1},\ n=2,3,\dots,\stoppingtime_0.$ Then
\begin{IEEEeqnarray}{rCl}
\stoppingtime &=& \sum_{n=1}^{\stoppingtime_0} \erasurecount_n.
\end{IEEEeqnarray}
Note that~$\curlybrac{\erasurecount_n}_{n=1}^{\infty}$ are~\iid~$\Bgeometricdist{1-\becerasureprob}$--distributed. Using Wald's identity~\cite[Eq. (84)]{wald1944cumulativesum} we conclude that
\begin{IEEEeqnarray}{rCl}
\Bexpectation{\stoppingtime} &=&  \Bexpectation{\stoppingtime_0} \Bexpectation{\bG_1}
\end{IEEEeqnarray}
from which~\eqref{expression_reduce_to_zero_erasure} follows.
\end{IEEEproof}\ignorespaces
\fi\ignorespaces
\section{Proof of Theorem~\ref{theorem_bec_achievability_iid_no_union_bound}}\label{appendix_proof_bec_achievability_iid_no_union_bound}
We consider the VLSF codebook ensemble specified by the set of all binary matrices with~$\messagecount$ rows and infinitely many columns. Furthermore, we assign a probability distribution on this ensemble by assuming each entry in the codebook being~\iid~$\Bbernoullidist{0.5}.$ Using the stopping time~\eqref{expression_bec_vlsf_optimal_stopping_time} and the decoder~\eqref{expression_bec_vlsf_optimal_decoder}, we can now create a VLSF code ensemble. By\iflongversion~Lemma~\ref{lemma_reduce_to_zero_erasure}\else~\cite[Lemma 10]{devassy2016nonasymptoticlong}\fi, we can write the ensemble average blocklength as
\begin{IEEEeqnarray}{rCl}
\Bexpectation{\stoppingtime}&=&\frac{\Bexpectation{\stoppingtime_0}}{1-\becerasureprob}\label{step_reduce_to_zero_erasure_iid_non_union_bound}
\end{IEEEeqnarray}
where $\stoppingtime_0$ is the stopping time when~$\becerasureprob=0.$ Let~$\messagebitsequalevent{w}{\chinp_1^n}$ be the event that the bits~$\chinp_1^n,$ which are distributed~\iid~$\Bbernoullidist{0.5},$ coincide with the first~$n$ bits of the codeword corresponding to message~$\Rinpmessage.$ Without loss of generality, we assume that message~$1$ is transmitted. The ensemble average of~$\stoppingtime_0$ is given by
\begin{IEEEeqnarray}{rCl}
\Bexpectation{\stoppingtime_0} &=&\iflongversion \sum_{n=0}^{\infty} \probof{\stoppingtime_0>n}\\
&=&\fi 1+\sum_{n=1}^{\infty} \probof{\bigcup_{w=2}^\messagecount \messagebitsequalevent{w}{\chinp_1^n}}\\
&=& \sum_{n=0}^{\infty} \parantheses{1-\parantheses{1-2^{-n}}^{\messagecount-1}}\\
&=& 1-\sum_{i=1}^{\messagecount-1}\binom{\messagecount-1}{i}\frac{\parantheses{-1}^i}{2^i-1}.\label{step_apply_non_union_bound_sum}
\end{IEEEeqnarray}
In~\eqref{step_apply_non_union_bound_sum}, we used the binomial theorem and the summation formula for geometric series. Substituting~\eqref{step_apply_non_union_bound_sum} into~\eqref{step_reduce_to_zero_erasure_iid_non_union_bound} we conclude that
\begin{IEEEeqnarray}{rCl}
\Bexpectation{\stoppingtime}&=&\frac{1}{1-\becerasureprob}\parantheses{1-\sum_{i=1}^{\messagecount-1}\binom{\messagecount-1}{i}\frac{\parantheses{-1}^i}{2^i-1}}.
\end{IEEEeqnarray}
Since there exists at least one VLSF code in the ensemble with average blocklength lower than the ensemble average blocklength, we conclude that~\eqref{expression_bec_achievability_iid_no_union_bound} must hold.
\iflongversion
\section{Proof of Theorem~\ref{theorem_bec_achiveability_vlsf_zero_error}}\label{appendix_proof_bec_achiveability_vlsf_zero_error}
We use the same steps as in the proof of Theorem~\ref{theorem_bec_achievability_iid_no_union_bound} except that the codebook ensemble is different. The columns are~\iid and uniformly distributed over the set of~$\messagecount$ dimensional vectors with~$\ceil{\messagecount/2}$ zeros. Using the same notation as in Appendix~\ref{appendix_proof_bec_achievability_iid_no_union_bound}, we have
\begin{IEEEeqnarray}{rCl}
\Bexpectation{\stoppingtime_0} &=& 1+\sum_{n=1}^{\infty} \probof{\bigcup_{w=2}^\messagecount \messagebitsequalevent{w}{\chinp_1^n}}\\
&\leq& \sum_{n=0}^{\infty} \min\parantheses{(\messagecount-1)\invprobofunequalbits^{-n},1}\label{step_apply_union_bound}\\
&=& \floor{m}+1+\frac{\invprobofunequalbits^{m-\floor{m}}}{\invprobofunequalbits-1}.\label{step_apply_geometric}
\end{IEEEeqnarray}
In~\eqref{step_apply_union_bound} we used the truncated union bound,~$\invprobofunequalbits$ is defined in~\eqref{expression_inv_prob},~$m=\log_\invprobofunequalbits\parantheses{\messagecount-1},$ and~\eqref{step_apply_geometric} follows from the summation formula for geometric series. Substituting~\eqref{step_apply_union_bound} into~\eqref{step_reduce_to_zero_erasure_iid_non_union_bound} we obtain~\eqref{expression_bec_achiveability_vlsf_zero_error}.

\fi\iflongversion
\section{Proof of Theorem~\ref{theorem_bec_achievability_expurgated_linear}}\label{appendix_proof_bec_achievability_expurgated_linear}
The proof follows again along the same lines as the proof of Theorem~\ref{theorem_bec_achievability_iid_no_union_bound}. This time, the ensemble contains only linear codes and is obtained as follows. The columns of the generator matrix  are independent and uniformly distributed over the set of all nonzero vectors in~$\binaryfinitefield^\messagebitcount.$ Since the code is linear, the decoder stops when the columns of the generator matrix corresponding to unerased positions form a basis for~$\binaryfinitefield^\messagebitcount.$ Then,~$\Bexpectation{\stoppingtime_0}$ can be interpreted as the average number of columns that need to be collected to obtain a basis for~$\binaryfinitefield^\messagebitcount.$ The dimension of the subspace of the first~$n$ columns of the random generator matrix can be modeled as a Markov chain with a single absorbing state (the state corresponding to maximum dimension~$\messagebitcount$). The time to absorption for this Markov chain follows a discrete phase-type distribution~\cite[Ch. 2]{neuts1981matrix}. Its expectation can be shown to be
\begin{IEEEeqnarray}{rCl}
\Bexpectation{\stoppingtime_0}&=& \messagebitcount+\sum_{i=1}^{\messagebitcount-1}\frac{2^{i}-1}{2^k-2^{i}}. \label{expression_avg_latency_to_full_rank}
\end{IEEEeqnarray}
We obtain~\eqref{expression_bec_achievability_expurgated_linear} by substituting~\eqref{expression_avg_latency_to_full_rank} into~\eqref{step_reduce_to_zero_erasure_iid_non_union_bound}.
\fi 
\end{appendices}

\bibliographystyle{IEEEtran}
\bibliography{IEEEabrv,confs-jrnls,publishers,refs}

\begin{thebibliography}{10}
\providecommand{\url}[1]{#1}
\csname url@samestyle\endcsname
\providecommand{\newblock}{\relax}
\providecommand{\bibinfo}[2]{#2}
\providecommand{\BIBentrySTDinterwordspacing}{\spaceskip=0pt\relax}
\providecommand{\BIBentryALTinterwordstretchfactor}{4}
\providecommand{\BIBentryALTinterwordspacing}{\spaceskip=\fontdimen2\font plus
\BIBentryALTinterwordstretchfactor\fontdimen3\font minus
  \fontdimen4\font\relax}
\providecommand{\BIBforeignlanguage}[2]{{%
\expandafter\ifx\csname l@#1\endcsname\relax
\typeout{** WARNING: IEEEtran.bst: No hyphenation pattern has been}%
\typeout{** loaded for the language `#1'. Using the pattern for}%
\typeout{** the default language instead.}%
\else
\language=\csname l@#1\endcsname
\fi
#2}}
\providecommand{\BIBdecl}{\relax}
\BIBdecl

\bibitem{ZeroErrorCapacityShanon}
C.~Shannon, ``The zero error capacity of a noisy channel,'' \emph{IRE Trans.
  Info. Theory}, vol.~2, no.~3, pp. 8--19, Sep. 1956.

\bibitem{doburshin1962anasymptotic}
R.~L. Dobrushin, ``An asymptotic bound for the probability error of information
  transmission through a channel without memory using the feedback,'' vol.~8,
  pp. 161--168, 1962.

\bibitem{burnashev1976data}
M.~V. Burnashev, ``Data transmission over a discrete channel with feedback.
  random transmission time,'' \emph{Probl. Inf. Transm.}, vol.~12, no.~4, pp.
  10--30, Dec. 1976.

\bibitem{YamamotoAchievability1979}
H.~Yamamoto and K.~Itoh, ``Asymptotic performance of a modified
  {S}chalkwijk-{B}arron scheme for channels with noiseless feedback,''
  \emph{{IEEE} Trans. Inf. Theory}, vol.~25, no.~6, pp. 729--733, Nov. 1979.

\bibitem{BerlinSimpleConverse2009}
P.~Berlin, B.~Nakibo{\u g}lu, B.~Rimoldi, and I.~Telatar, ``A simple converse
  of {Burnashev's} reliability function,'' \emph{{IEEE} Trans. Inf. Theory},
  vol.~55, no.~7, pp. 3074--3080, Jul. 2009.

\bibitem{JensenShannonDivergenceNaghshvar2015}
M.~Naghshvar, T.~Javidi, and M.~Wigger, ``Extrinsic {Jensen}-{Shannon}
  divergence: Applications to variable-length coding,'' \emph{{IEEE} Trans.
  Inf. Theory}, vol.~61, no.~4, pp. 2148--2164, Apr. 2015.

\bibitem{polyanskiy2011feedback}
Y.~Polyanskiy, H.~V. Poor, and S.~Verd{\'u}, ``Feedback in the non-asymptotic
  regime,'' \emph{{IEEE} Trans. Inf. Theory}, vol.~57, no.~8, pp. 4903--4925,
  Aug. 2011.

\bibitem{polyanskiy2010channel}
------, ``Channel coding rate in the finite blocklength regime,'' \emph{{IEEE}
  Trans. Inf. Theory}, vol.~56, no.~5, pp. 2307--2359, May 2010.

\bibitem{ForneyExponentialBounds1968}
G.~D. Forney~Jr, ``Exponential error bounds for erasure, list, and decision
  feedback schemes,'' \emph{{IEEE} Trans. Inf. Theory}, vol.~14, no.~2, pp.
  206--220, Mar. 1968.

\bibitem{mackay2005fountain}
D.~J. MacKay, ``Fountain codes,'' \emph{Inst. Electr. Eng. Proc.-Commun.}, vol.
  152, no.~6, pp. 1062--1068, Dec. 2005.

\bibitem{massay2007zeroerror}
J.~L. Massey, ``Zero error,'' presented at the IEEE Winter School on Coding and
  Information Theory, Mar 2007.

\bibitem{eggleston1958convexity}
H.~G. Eggleston, \emph{Convexity}.\hskip 1em plus 0.5em minus 0.4em\relax New
  York: Cambridge university press, 1958.

\bibitem{wald1945sequential}
A.~Wald, ``Sequential tests of statistical hypotheses,'' \emph{Ann. Math.
  Statist.}, vol.~16, no.~2, pp. 117--186, Jun. 1945.

\bibitem{tartakovsky2014sequential}
A.~Tartakovsky, I.~Nikiforov, and M.~Basseville, \emph{Sequential analysis:
  Hypothesis testing and changepoint detection}.\hskip 1em plus 0.5em minus
  0.4em\relax Boca Raton, Florida, USA: CRC Press, 2014.

\bibitem{ShannonGallagher1967522}
C.~Shannon, R.~Gallager, and E.~Berlekamp, ``Lower bounds to error probability
  for coding on discrete memoryless channels. \{II\},'' \emph{Information and
  Control}, vol.~10, no.~5, pp. 522 -- 552, Jan. 1967.

\bibitem{neuts1981matrix}
M.~Neuts, \emph{Matrix-geometric Solutions in Stochastic Models: An Algorithmic
  Approach}.\hskip 1em plus 0.5em minus 0.4em\relax Baltimore, MD: The Johns
  Hopkins Univ. Press, 1981.

\bibitem{huffman1952method}
D.~A. Huffman, ``A method for the construction of minimum redundancy codes,''
  \emph{IRE Trans. Info. Theory}, vol.~40, no.~9, pp. 1098--1101, Sep. 1952.

\bibitem{ahlswede2006general}
R.~Ahlswede, ``Identification entropy,'' in \emph{General Theory of Information
  Transfer and Combinatorics}.\hskip 1em plus 0.5em minus 0.4em\relax Berlin,
  Germany: Springer-Verlag, 2006, vol. 4123, pp. 595--613.

\bibitem{Burkholder1963ptimumProperties}
D.~L. Burkholder and R.~A. Wijsman, ``\BIBforeignlanguage{English}{Optimum
  properties and admissibility of sequential tests},''
  \emph{\BIBforeignlanguage{English}{Ann. Math. Statist.}}, vol.~34, no.~1, pp.
  1--17, Mar. 1963.

\bibitem{Wald1948OptimumCharacter}
A.~Wald and J.~Wolfowitz, ``\BIBforeignlanguage{English}{Optimum character of
  the sequential probability ratio test},''
  \emph{\BIBforeignlanguage{English}{Ann. Math. Statist.}}, vol.~19, no.~3, pp.
  326--339, Sep. 1948.

\bibitem{wald1944cumulativesum}
A.~Wald, ``On cumulative sums of random variables,'' \emph{Ann. Math.
  Statist.}, vol.~15, no.~3, pp. 283--296, Sep. 1944.

\end{thebibliography}

\end{document}